\documentclass[12 pt, article]{article}
\usepackage[left=1in, right=1in, top=0.75in, bottom=0.75in]{geometry}
\usepackage[utf8]{inputenc}
\usepackage{setspace}
\usepackage{graphicx}
\usepackage{authblk}
\usepackage{sectsty}
\usepackage{gensymb}
\usepackage{parskip}
\usepackage{caption}
\usepackage{titlesec}
\usepackage{wrapfig}
\usepackage{xcolor}
\usepackage[utf8]{inputenc}
\usepackage{amsmath}
\graphicspath{{images/}}
\sectionfont{\fontsize{13}{16}\selectfont}
\subsectionfont{\fontsize{12}{12}\selectfont}

\titlespacing\section{0pt}{12pt plus 2pt minus 2pt}{12pt plus 2pt minus 2pt}
\usepackage[backend=biber, style=nature]{biblatex}
\title{\Large\textbf {Atomic engineering of interfacial polarization switching in van der Waals multilayers}}

\author[1]{Madeline Van Winkle}
\author[1]{Nikita Dowlatshahi}
\author[1]{Nikta Khaloo}
\author[1,2]{Mrinalni Iyer}
\author[1]{Isaac M. Craig}
\author[3]{Rohan Dhall}
\author[4]{Takashi Taniguchi}
\author[5]{Kenji Watanabe}
\author[1,6,*]{D. Kwabena Bediako}
\affil[1]{\textit{Department of Chemistry, University of California, Berkeley, CA 94720, USA}}
\affil[2]{\textit{Department of Chemistry, University of Minnesota, Minneapolis, MN 55455, USA}}
\affil[3]{\textit{Molecular Foundry, Lawrence Berkeley National Laboratory, Berkeley, CA 94720, USA}}
\affil[4]{\textit{International Center for Materials Nanoarchitectonics, National Institute for Materials Science, 1-1 Namiki, Tsukuba 305-0044, Japan}}
\affil[5]{\textit{Research for Functional Materials, National Institute for Materials Science, 1-1 Namiki, Tsukuba 305-0044, Japan}}
\affil[6]{\textit{Chemical Sciences Division, Lawrence Berkeley National Laboratory, Berkeley, CA 94720, USA}}
\affil[*]{Correspondence to: bediako@berkeley.edu}

\date{}
\begin{document}
\maketitle

\doublespacing
\textbf{In conventional ferroelectric materials, polarization is an intrinsic property limited by bulk crystallographic structure and symmetry. Recently, it has been demonstrated that polar order can also be accessed using inherently non-polar van der Waals materials through layer-by-layer assembly into heterostructures, wherein interfacial interactions can generate spontaneous, switchable polarization. Here, we show that introducing interlayer rotations in multilayer vdW heterostructures modulates both the spatial ordering and switching dynamics of polar domains, engendering unique tunability that is unparalleled in conventional bulk ferroelectrics or polar bilayers. Using operando transmission electron microscopy we show how changing the relative rotations of three \(\mathbf{WSe_2}\) layers produces structural polytypes with distinct arrangements of polar domains, leading to either a global or localized switching response. Introducing uniaxial strain generates structural anisotropy that yields a range of switching behaviors, coercivities, and even tunable biased responses. We also provide evidence of physical coupling between the two interfaces of the trilayer, a key consideration for controlling switching dynamics in polar multilayer structures more broadly.}

\newpage
Ferroelectric (FE) materials with three-dimensional lattices have been employed in an array of applications, including nonvolatile memory, actuators, and sensors, for decades.\supercite{haertling1999ferroelectric,mikolajick2020past} 
However, further miniaturization of electronic devices relies on the realization and manipulation of polar order in atomically thin crystals, such as two-dimensional van der Waals (vdW) materials. While many bulk, layered vdW crystals are naturally centrosymmetric and therefore non-polar, layer-by-layer assembly of individual vdW layers has been used to build non-centrosymmetric 2D heterostructures possessing an out-of-plane, interfacial polarization that can be switched via sliding of one layer. This bottom-up approach greatly expands the number of potential 2D FE candidates and has been demonstrated in various common vdW materials, including hexagonal boron nitride (hBN)\supercite{yasuda2021stacking,vizner2021interfacial} and transition metal dichalcogenides (TMDs).\supercite{wang2022interfacial,weston2022interfacial,rogee2022ferroelectricity,deb2022cumulative,meng2022sliding,ko2023operando}

Much of the work in the field thus far has centered on polar bilayer heterostructures; however, recently it was demonstrated that interfacial polarization is cumulative in multilayer systems, enabling access to a ladder of polarization states depending on the number of layers and the translational offsets between them.\supercite{deb2022cumulative,meng2022sliding} In this study, we use operando transmission electron microscopy (TEM) to show how controlling the relative rotations between atomic layers in twisted trilayer tungsten diselenide (TTL-\(\mathrm{WSe_2}\)) dictates the arrangement of polar domains in the resulting structure, offering control over a global versus local switching response. We also observe that uniaxial strain engineers a range of switching dynamics, including moiré anti-ferroelectric, ferroelectric, and distinctively, biased responses. Further, important interactions between the two interfaces of the trilayer are observed through coupling between intralayer strain and consequent switching behavior as well as pinning between commensurately stacked domain walls, providing insight into the cooperative nature of polarization switching in multilayer heterostructures and offering new routes to electrical control over interlayer twist angle in moiré superlattices.

%%%%%%%%%%%%%%%%%%%%%%%%%%%%%%%%%%%%%%%%%%%%%%%%
%% FIGURE 1 
%%%%%%%%%%%%%%%%%%%%%%%%%%%%%%%%%%%%%%%%%%%%%%%%
\begin{figure*}[htbp]
    \centerline{\includegraphics[width=\textwidth]{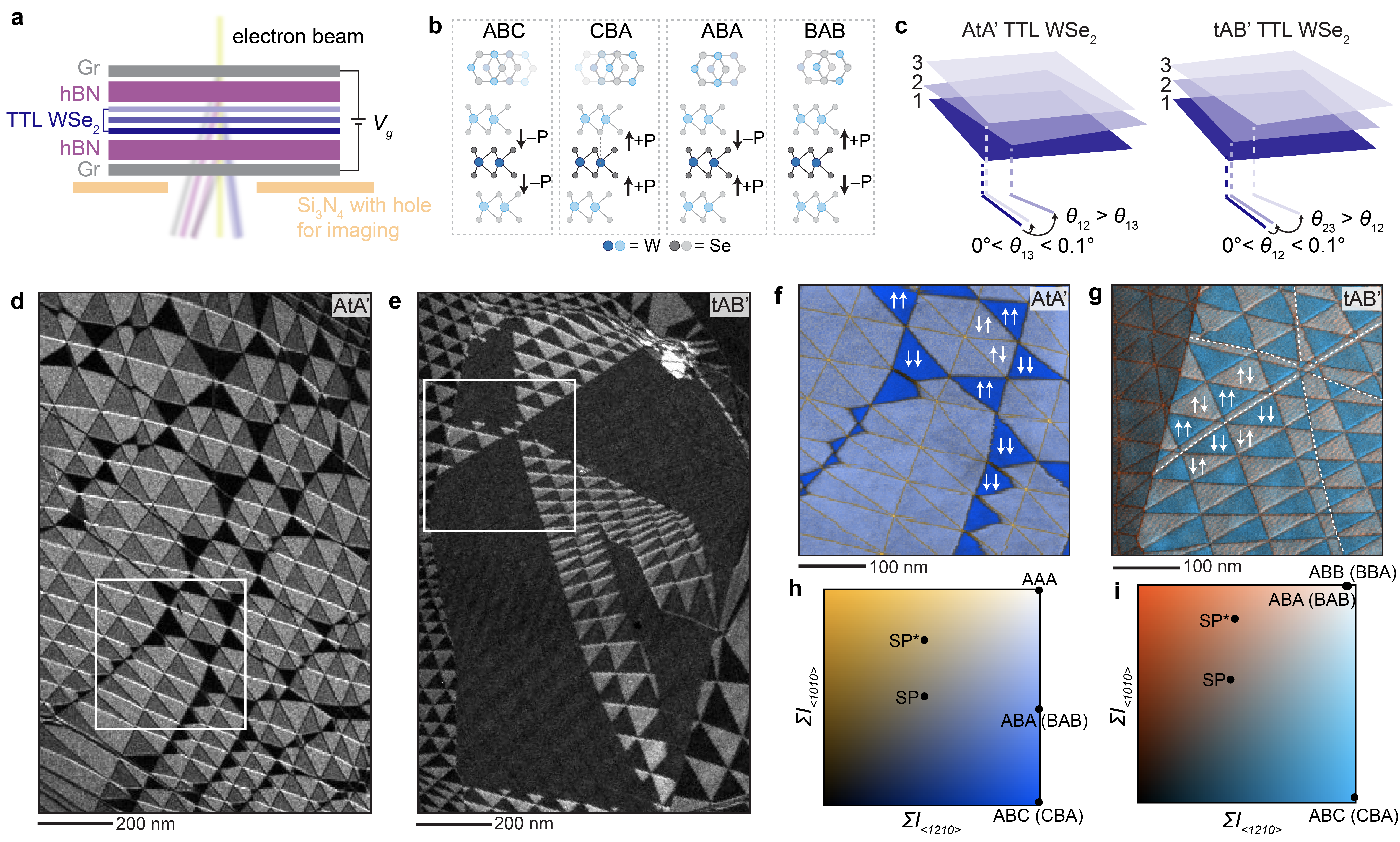}}
    \caption{\textbf{Twisted trilayer \({\mathbf{WSe_2}}\) polytypes and assignment of polar domains. a,} Schematic of twisted trilayer (TTL) \(\mathrm{WSe_2}\) device on a silicon nitride (\(\mathrm{Si_3N_4}\)) membrane for operando TEM studies. Hexagonal boron nitride, hBN. Graphite, Gr.\textbf{b,} Low-energy atomic stacking configurations for parallel-stacked trilayer \(\mathrm{WSe_2}\), with the out-of-plane polarization direction (+P or –P) at each interface indicated by arrows. ABC and CBA configurations have equal and opposite non-zero net polarizations, while ABA and BAB configurations have no net polarization. \textbf{c,} Illustrations of two main TTL polytypes, the A-twist-A\(^\prime\) (AtA\(^\prime\)) type and twist-A-B\(^\prime\) (tAB\(^\prime\)) type. \textbf{d,e,} Dark-field (DF) TEM images of AtA\(^\prime\) and tAB\(^\prime\) \(\mathrm{WSe_2}\) devices, respectively, obtained using the [10\(\bar{1}\)0] Bragg reflection. \textbf{f,g,} Four-dimensional STEM (4D-STEM) virtual DF images from boxed regions in \textbf{d,e}. Arrows indicate the local direction of polarization at the two trilayer interfaces. \textbf{h,i,} Color legends for virtual DF images in \textbf{f,g}, respectively, relating pixel color to the normalized cumulative first and second order Bragg disk intensities (\(\Sigma I_{\langle{1010}\rangle}\) and \(\Sigma I_{\langle{1210}\rangle}\)) obtained from the local experimental convergent beam electron diffraction pattern. Plotted points indicate normalized values calculated for high-symmetry stacking configurations using multislice simulations.}
\end{figure*}

\section*{Polar domain order in twisted trilayer \(\mathbf{WSe_2}\) polytypes}
Samples were prepared using a dry transfer method and consist of the twisted trilayer (stacked with a parallel orientation between \(\mathrm{WSe_2}\) layers) sandwiched between thin hBN dielectric sheets (each \(<15\) nm thick) and few-layer graphite electrodes (see Methods and Supplementary Section 1). The full heterostructure is thin enough for transmission of an incident electron beam, enabling TEM imaging of the device while simultaneously applying an out-of-plane electric field across the sample (Fig. 1a). Due to the presence of a rotational offset between the \(\mathrm{WSe_2}\) layers, there is a spatial variation in the crystallographic stacking order, referred to as a moiré superlattice, throughout the trilayer; however, as a consequence of spontaneous relaxation of this superlattice,\supercite{yoo2019atomic,weston2020atomic,craig2023local} four configurations with the lowest stacking energy comprise the majority of the moiré structure. Of these, the ABC and CBA stacking types have equal and opposite non-zero net polarization due to uncompensated charge transfer at the two stacked interfaces, while the ABA and BAB stackings have no net polarization due to mirror symmetry (Fig. 1b). The spatial arrangement of polar and non-polar domains depends on the relative rotations between the \(\mathrm{WSe_2}\) sheets. In the A-twist-A\(^\prime\) (AtA\(^\prime\)) polytype, the outer layers (layers 1 and 3) are nearly aligned (\(0\degree<\theta_{13}<0.1\degree\)) while the middle layer (layer 2) is rotated further (\(\theta_{12} > \theta_{13}\)). In the twist-A-B\(^\prime\) (tAB\(^\prime\)) polytype the bottom and middle layers are nearly aligned (\(0\degree<\theta_{12}<0.1\degree\)) while the top layer is rotated further (\(\theta_{23}>\theta_{12}\)) (Fig. 1c). 

To view the moiré structure in these two polytypes we use dark-field TEM (DF-TEM), which is a diffraction-based imaging technique that can be used to filter signal in multi-component structures,\supercite{huang2011grains,yoo2019atomic} making it well-suited for the observation of buried interfaces. As shown in the DF-TEM images in Fig. 1d,e, the AtA\(^\prime\) and tAB\(^\prime\) polytypes produce distinct moiré patterns. In the AtA\(^\prime\) polytype we observe a kagome-like pattern superimposed on a triangular superlattice. Since \(\theta_{12}>\theta_{13}\), the longer lengthscale kagome structure is attributed to the moiré between the outer layers and the shorter triangular structure to the moiré between the outer layers and middle layer. In the tAB\(^\prime\) polytype there are two superimposed triangular superlattices; the superlattice with the longer (shorter) periodicity comes from the bottom and middle (middle and top) layers since \(\theta_{23}>\theta_{12}\). The diffraction contrast observed in the DF-TEM images arises from differences in local stacking and from the presence of a global tilt of the sample away from the zone axis (Supplementary Section 2).\supercite{sung2019stacking} 

To identify which regions of the observed moiré structures have a net polarization, we use interferometric four-dimensional scanning transmission electron microscopy (4D-STEM, Fig. 1f,g). This imaging technique relates the intensities in interfering electron diffraction disks to the local translational offset between layers (Supplementary Section 3) and therefore can be used to map out the stacking order throughout the moiré.\supercite{kazmierczak2021strain,zachman2021interferometric,van2023rotational,craig2023local} We first perform multislice simulations to calculate the cumulative diffraction intensity for the \(\langle1010\rangle\) and \(\langle1210\rangle\) Bragg peaks for the high-symmetry stacking orders in both polytypes (Fig. 1h,i). These computed values are then compared to the experimental 4D-STEM diffraction intensities (Supplementary Section 3) to assign the stacking order in each domain in the scan region. In the AtA\(^\prime\) polytype, we find that the ABC and CBA (i.e., polar) domains are localized in the kagome-like structures, alternating in the points of each star, and on either side of elongated domain walls, the origin of which will be discussed later. Meanwhile the tAB\(^\prime\) polytype is divided into large sections containing either ABC or CBA stacking domains in periodic triangular arrangements. We note that interferometric 4D-STEM is purely a structural diagnostic and therefore can not be used to measure the local polarization (P) direction (i.e., net +P versus -P); however, the polarization direction in each domain can be confirmed by observing the response to an applied electric field, which we discuss next.

\section*{Polarization switching in AtA\(^\prime\) polytype}
%%%%%%%%%%%%%%%%%%%%%%%%%%%%%%%%%%%%%%%%%%%%%%%%
%% FIGURE 2
%%%%%%%%%%%%%%%%%%%%%%%%%%%%%%%%%%%%%%%%%%%%%%%%
\begin{figure*}[htbp]
    \centerline{\includegraphics[width=160mm]{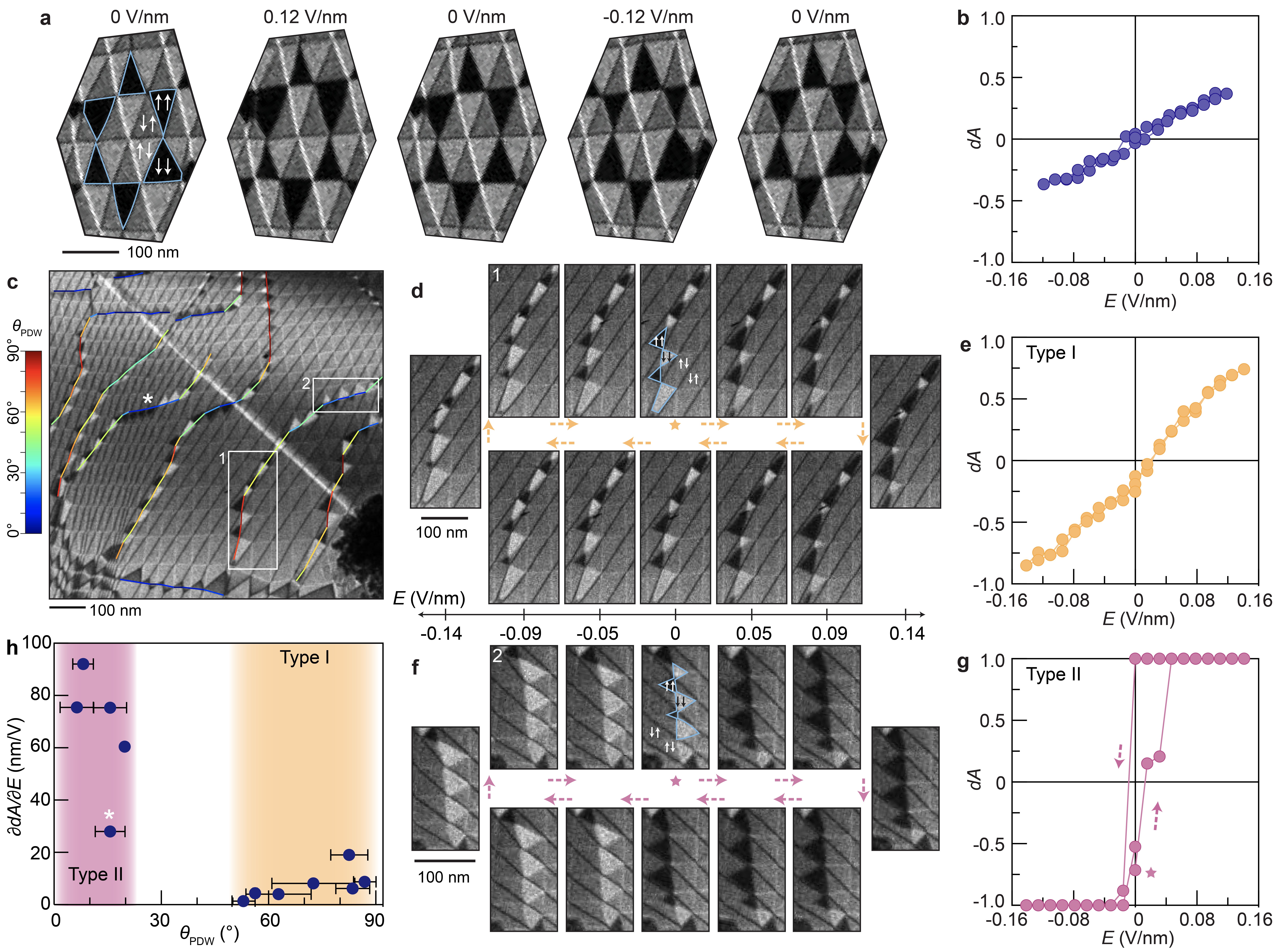}}
    \caption{\textbf{Polar domain dynamics in AtA\(^\prime\) \(\mathbf{WSe_2}\). a,} Sample DF-TEM images and \textbf{b,} corresponding plot of normalized net polarization (\textit{dA}) as a function of applied field (\textit{E}) for a minimally heterostrained AtA\(^\prime\) kagome-like structure. \textbf{c,} Low-magnification DF-TEM image of a slightly heterostrained (\(<0.03\%\)) AtA\(^\prime\) \(\mathrm{WSe_2}\) sample with polar domain walls (PDW). PDW orientations (\(\theta_{PDW}\)) overlaid in color. White numbered boxes highlight regions analyzed in \textbf{d–g}. \textbf{d–g,} Sample DF-TEM images and corresponding plots of \textit{dA} versus \textit{E} for PDWs with a (\textbf{d,e}) gradual and (\textbf{f,g}) sharp switching response. Colored stars mark initial positions and colored dashed arrows show scan directions. \textbf{h,} PDW polarizability (\(\delta dA/\delta E\)) versus orientation (\(\theta_{PDW}\)). Shading highlights two groups of PDWs, one with gradual (Type I) and one with sharp (Type II) switching. Asterisks in \textbf{c,h} mark PDW with sequential switching, analyzed further in Fig. 3. White and black arrows in \textbf{a,d,f} indicate local polarization at the two trilayer interfaces. Light blue lines in \textbf{a,d,f} mark regions used for analysis in \textbf{b,e,g}, respectively. All DF-TEM images produced from [10\(\bar{1}\)0] Bragg reflection. Differences in sample tilts cause variability in intensities of ABC and CBA domains.}
\end{figure*}

The distinct arrangements of polar domains in the AtA\(^\prime\) and tAB\(^\prime\) trilayers suggest that these two polytypes will behave differently in an applied field. We first investigate the structural response in the AtA\(^\prime\) polytype. Fig. 2a shows how one of the star structures from the sample in Fig. 1d distorts under application of an electric bias. At zero field, the ABC (-P) and CBA (+P) regions are approximately the same size because they are energetically degenerate. As a negative (positive) bias is applied, this degeneracy is lifted and the ABC (CBA) regions grow while the CBA (ABC) regions shrink (Supplementary Video 1). To quantify this change, we define an order parameter, \textit{dA}, which is the normalized relative area of the ABC and CBA domains, described by Eq.1:

\begin{equation}
    dA = \frac{A_{CBA} - A_{ABC}} {A_{CBA} + A_{ABC}}
\end{equation}

Previous studies have demonstrated that this order parameter can be used as a proxy for net polarization.\supercite{ko2023operando} The variation in \textit{dA} in the star structure is plotted as a function of field (\textit{E}) in Fig. 2b, showing an approximately linear relationship with minimal structural hysteresis or remnant net polarization. In polar twisted bilayers, this behavior has been referred to as a moiré anti-ferroelectric response (MAFE)\supercite{ko2023operando} and is a direct consequence of the moiré structure, wherein the topology of the moiré domain wall network precludes complete switching of polar domains.\supercite{alden2013strain,enaldiev2022scalable,engelke2023topological} We do not observe structural distortions in the surrounding non-polar regions (Supplementary Section 4) except as a consequence of deformation of nearby polar domains, such as in the center of the star in Fig. 2a. 

It is also common for moiré structures to acquire heterostrain during sample fabrication, where one or more layers are uniaxially stretched relative to the adjacent layer(s).\supercite{huder2018electronic, edelberg2020tunable, lau2022reproducibility}  Moiré structures with a large periodicity are particularly susceptible to substantial distortions even in the presence of a small atomic scale heterostrain (Supplementary Section 5).\supercite{cosma2014moire} A DF-TEM image of an AtA\(^\prime\) sample with a very slight misalignnment between the outer layers (\(\theta_{13}<0.03\degree\)) and an estimated heterostrain \(<0.03\%\) (Supplementary Fig. 9) is shown in Fig. 2c. This sample contains extended polar domain walls (PDWs) in the longer lengthscale moiré rather than a kagome-like pattern. Similar elongated domain walls have been observed in heterostrained twisted trilayer graphene.\supercite{craig2023local} Interestingly, in this heterostrained sample we observe two groups of PDWs with different responses to an applied field (Supplementary Videos 2–4). The first type displays a gradual response to the field (Fig. 2d,e) where the ABC and CBA domains steadily grow and shrink during biasing. However, the second type has a much sharper response where the polar domains rapidly flip between ABC and CBA stacking (Fig. 2f,g), similar to an untwisted polar heterostructure but on a much shorter lengthscale (10s rather than 100s of nm). Additionally, this second domain wall type has structural hysteresis and a remnant net polarization at zero field, a FE response. In the PDW shown in Fig. 2f,g, the coercivity is relatively small due to the small domain size; however some FE PDWs, discussed in the next section, exhibit a larger coercivity. Uniquely, the polar domains in the heterostrained twisted trilayer structures shown here are not confined by the same topology as either a twisted bilayer or a non-heterostrained twisted trilayer, enabling a true ferroelectric response in a moiré material for the first time. 

To understand the origin of the different switching responses exhibited by these two groups of domain walls, we calculate the polarizability for various PDW segments with different orientations for the sample shown in Fig. 2c. Here, polarizability, the ease with which the net polarization switches, is equivalent to the slope of \textit{dA} versus \textit{E} (\(\delta dA/\delta E\)). Fig. 2c illustrates the orientation of the PDWs in the sample (\(\theta_{PDW}\)) and Fig. 2h relates the polarizability of a selection of these PDWs to their average orientation (additional information in Supplementary section 6). Based on this investigation, it is evident that, for this sample, PDWs oriented around 60-90\(\degree\) relative to the \textit{x}-axis have a polarizability that is roughly 3-4 times lower than PDWs oriented closer to 0\(\degree\). Knowing that domain polarizability is affected by domain anisotropy in twisted bilayers,\supercite{molino2022ferroelectric,ko2023operando} the phenomenon we observe could be attributed to uniaxial heterostrain present in the superimposed small moiré, which is oriented on average around 72\(\degree\) relative to the \textit{x}-axis in this sample (Supplementary Fig. 10). This heterostrain introduces global structural anisotropy in the trilayer, as seen by the distortion of the triangular moiré unit cells, which could produce a preferential sliding axis for domain switching, analogous to an easy axis for spin reorientation in magnetic systems.\supercite{johnson1996magnetic,paes2013effective} Overall, it is clear that polarization switching in the AtA\(^\prime\) trilayer is influenced by the uniaxial strain fields in all three \(\mathrm{WSe_2}\) layers collectively.

\section*{Heterostrain control of polarization bias}
%%%%%%%%%%%%%%%%%%%%%%%%%%%%%%%%%%%%%%%%%%%%%%%%
%% FIGURE 3
%%%%%%%%%%%%%%%%%%%%%%%%%%%%%%%%%%%%%%%%%%%%%%%%
\begin{figure*}[htbp]
    \centerline{\includegraphics[width=156mm]{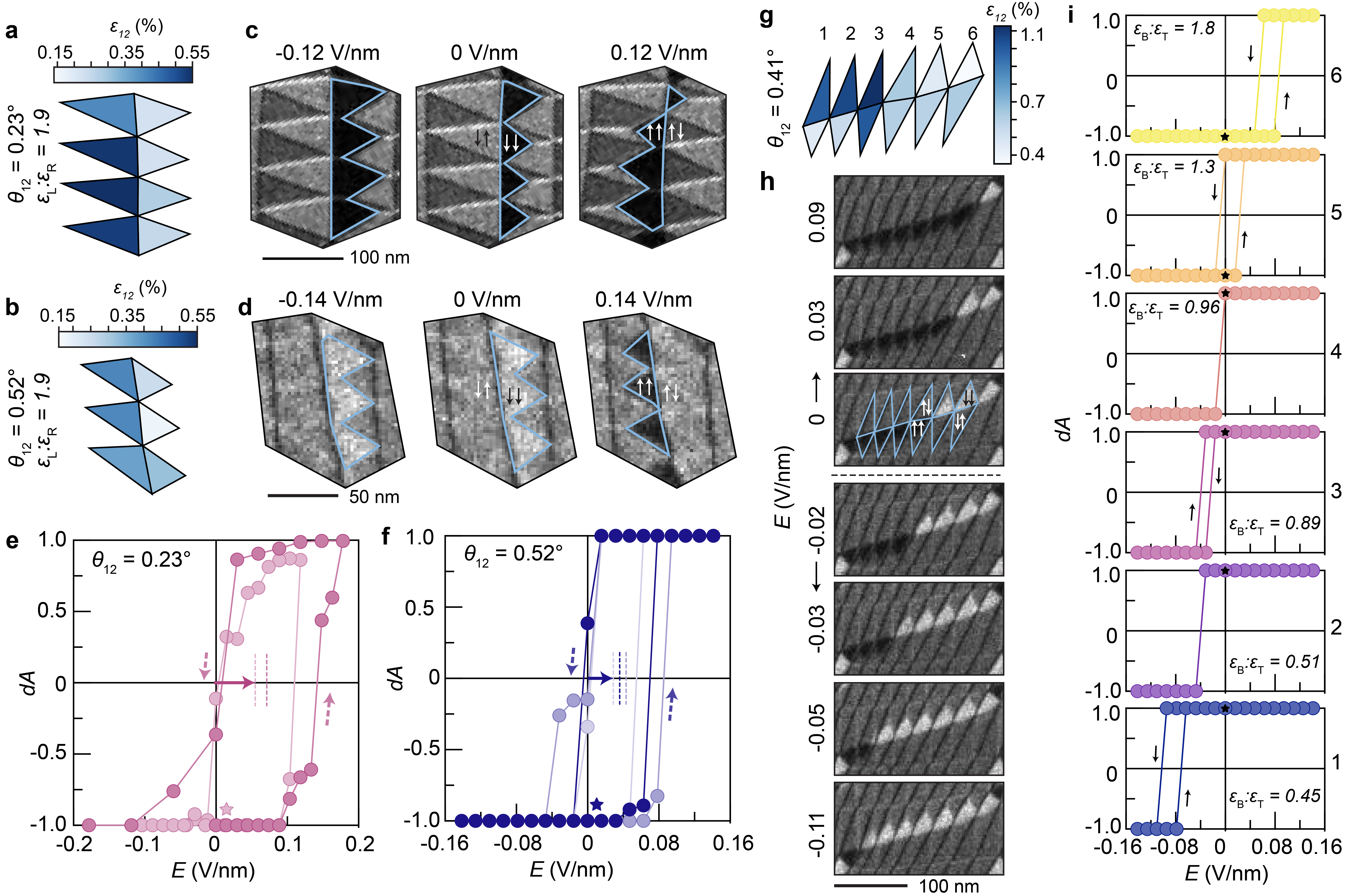}}
    \caption{\textbf{Biased polarization from heterostrain gradients in AtA\(^\prime\) polytype. a,b,} Maps of heterostrain in the smaller moiré (\(\epsilon_{12}\)), \textbf{c,d,} sample DF-TEM images at different applied fields, and \textbf{e,f,} plots of normalized net polarization (\textit{dA}) as a function of applied field (\textit{E}) for two PDWs, one with \(\theta_{12} = 0.23\degree\)(\textbf{a,c,e}) and the other with \(\theta_{12} = 0.52\degree\)(\textbf{b,d,f}). Vertical dashed lines in \textbf{e,f} highlight shift of the center of the hysteresis loop away from 0 V/nm for each biasing cycle pictured. \textbf{g,} Map of \(\epsilon_{12}\) and \textbf{h,} sample DF-TEM images at different applied fields for a PDW with a variation in \(\epsilon_{12}\) along the length of the domain wall (average \(\theta_{12} = 0.41\degree\)). \textbf{i,} Plots of \textit{dA} as a function \textit{E} for each of the six polar domains pictured in \textbf{g,h}. \(\epsilon_L:\epsilon_R\) and \(\epsilon_B:\epsilon_T\) indicate the heterostrain ratio between small moiré domains on either the left (L) and right (R) side (\textbf{a–d}) or bottom (B) and (T) (\textbf{g–i}) of the PDW. In \textbf{c,d,h} white and black arrows indicate polarization at the two interfaces of select domains and light blue lines mark polar regions used for the calculation of \textit{dA}. All DF-TEM images produced from the [10\(\bar{1}\)0] Bragg reflection.}
\end{figure*}

Some PDWs run through a heterostrain gradient where the small moiré is considerably more distorted on one side of the domain wall than the other. For example, in the regions shown in Fig. 3a,b, the magnitude of heterostrain between layers 1 and 2 (\(\epsilon_{12}\)) is approximately twice as large on the left side of the PDW compared to the right side (\(\epsilon_L:\epsilon_R = 1.9\)). In both regions, which have different interlayer twist angles between the outer and middle layers (\(\theta_{12}\)), we observe that the polar domains localize on the less heterostrained half of the PDW at zero field. It has been demonstrated that the presence of anisotropic strain in multilayer graphene further increases the energy of the rhombohedral (ABC/CBA) stacking types relative to the Bernal (ABA/BAB) stacking.\supercite{geisenhof2019anisotropic} Presuming that \(\mathrm{WSe_2}\) behaves similarly, we rationalize that the observed preferential alignment of polar domains at zero field arises from an intrinsic thermodynamic bias generated by the appreciable difference in heterostrain between the sides of the PDW. Plotting \textit{dA} versus \textit{E} (Fig. 3e,f) over two or three biasing cycles, we see that this energetic preference toward one polarization state at zero field ultimately shifts the hysteresis curve along the electric field axis. We note that the enhanced coercivity of these two PDWs compared to that shown in Fig. 2g likely stems from a combination of structural factors, such as domain size, local distortions of the small moiré after polarization switching, and random variations in structural pinning sites at the ends of the domain wall segments pictured. Nevertheless, it is evident that the heterostrain gradient effectively pins the polar domains along one scan direction to produce a consistently biased FE response.

Remarkably, this asymmetry in the polarization curve can be systematically tuned down to the length scale of an individual polar domain. To demonstrate, we consider a PDW in a continuously varying heterostrain gradient. Fig. 3g illustrates that \(\epsilon_{12}\) is smaller on the bottom side of the PDW than on the top for domains 1–4 and vice versa for domains 5–6. Based on this variation in the direction of the heterostrain gradient, at zero field polar domains 1–4 align on the bottom side of the PDW and polar domains 5–6 align on the top, as shown in Fig. 3h. As a field is applied, each domain responds independently in a domino-like fashion (Fig. 3h), where the field necessary for polarization switching is directly correlated to the magnitude and direction of the local heterostrain gradient (Fig. 3i). These results demonstrate that coupling strain engineering\supercite{lee2011giant,jeon2013flexoelectric,hou2022nonvolatile} with moiré engineering could offer a route for precise, spatially localized manipulation of interfacial polarization. 

%Such stabilization of a particular aligned state at zero field can be leveraged in nonvolatile memories and logic devices, as demonstrated with exchange-biased magnetic structures.\supercite{nogues1999exchange,gider1998magnetic, parkin1999exchange, skumryev2003beating} In conventional ferroelectrics, this phenomenon is referred to as an internal bias field and has most commonly been associated with mobile defects and structural aging or fatiguing,\supercite{genenko2015mechanisms} making it difficult to control. 

%Notably, this strain-induced pinning considerably increases the coercivities of the PDWs compared to those discussed in the previous section (see Fig. 1f,g) and can be tuned by the moiré structure. Namely, larger polar domains (produced by smaller values of \(\theta_{12}\)) appear to be more strongly pinned by the heterostrain gradient and are therefore more biased and more coercive (Fig. 3e) than smaller domains (Fig. 3f). 

\section*{Polar domain switching in tAB\(^\prime\) polytype}
%%%%%%%%%%%%%%%%%%%%%%%%%%%%%%%%%%%%%%%%%%%%%%%%
%% FIGURE 4
%%%%%%%%%%%%%%%%%%%%%%%%%%%%%%%%%%%%%%%%%%%%%%%%
\begin{figure*}[tbp]
    \centerline{\includegraphics[width=\textwidth]{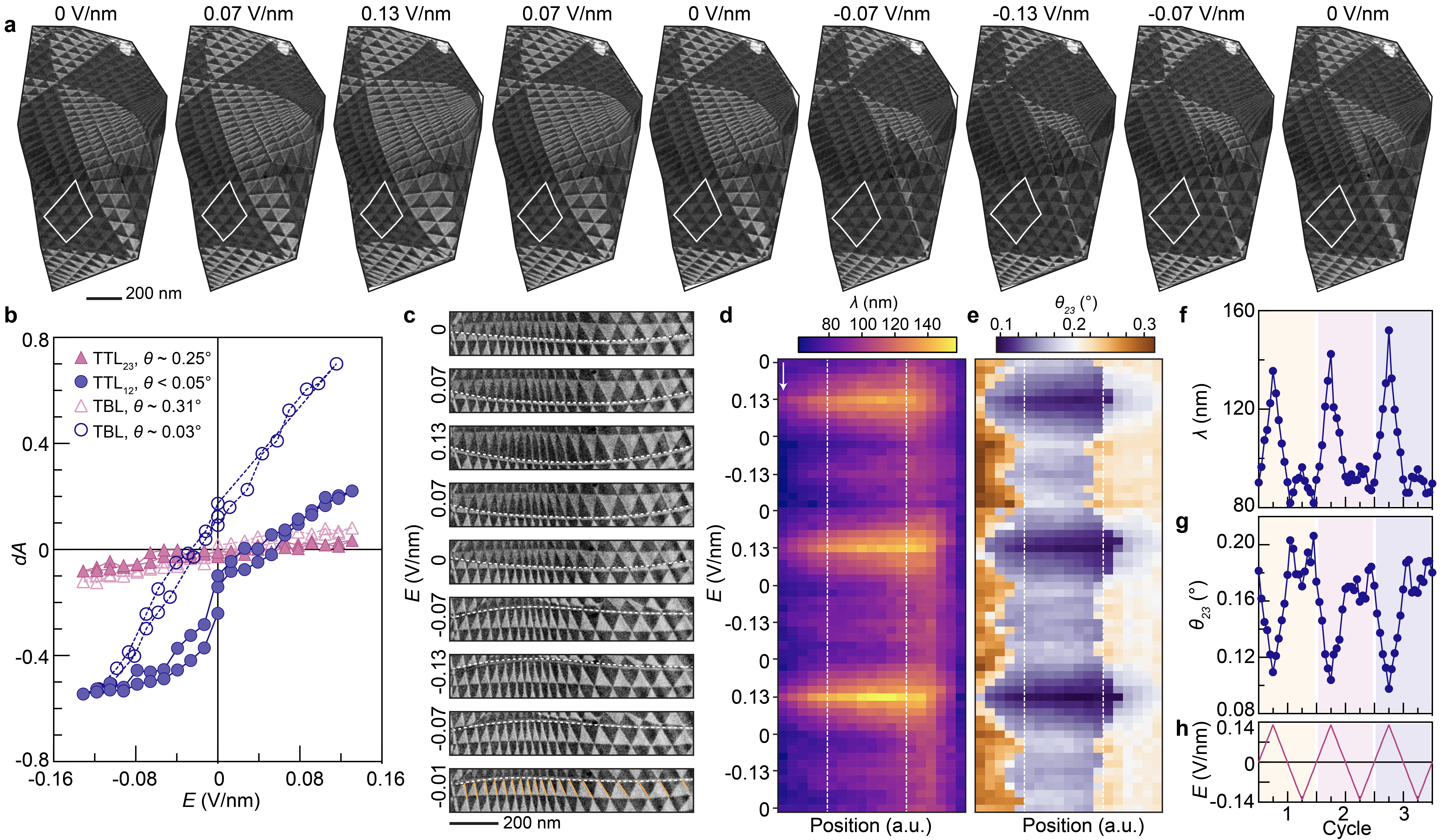}}
    \caption{\textbf{Polarization switching in tAB\(^\prime\) polytype. a,} DF-TEM images for a tAB\(^\prime\)-type \(\mathrm{WSe_2}\) trilayer over a series of applied fields, produced from the [10\(\bar{1}\)0] Bragg reflection. \textbf{b,} Normalized relative area of light versus dark domains (\textit{dA}) as a function of applied field (\textit{E}), describing deformation of the large and small moiré of the twisted trilayer (TTL) shown in \textbf{a} (\(\theta_{12}<0.05\degree\) and \(\theta_{23}=0.25\degree\)) compared to analagous twisted bilayer (TBL) structures (\(\theta\approx0.03\degree\) and \(\theta=0.31\degree\)). The area used for analysis of the small moiré in the TTL is boxed in white in \textbf{a}. The full sample area in \textbf{a} was used for analysis of the TTL large moiré. \textbf{c,} Magnified DF-TEM images showing pinning of a domain wall in the large TTL moiré to domain walls in the smaller moiré as a field is applied. Images produced from the [1\(\bar{1}\)00] Bragg reflection. \textbf{d,e,} Plots tracking evolution of side length (\(\lambda\)) and local twist angle (\(\theta_{23}\)) for a row of small moiré domains shown in \textbf{c} over three cycles of biasing. Side lengths used for \textbf{d} are marked with orange lines in \textbf{c}. \textbf{f,g,} Corresponding average values of \(\lambda\) and \(\theta_{23}\), respectively, over the course of three biasing cycles (demarcated by yellow, red, and blue shading and described by the applied field profile illustrated in \textbf{h}). Averages calculated from the section of \textbf{d} and \textbf{e} boxed in white.}
\end{figure*}

Next we discuss how the tAB\(^\prime\) polytype responds to an applied field. Whereas the AtA\(^\prime\) polytype had a highly localized polarization switching response, the tAB\(^\prime\) structure globally deforms under application of an electric field (Fig. 4a, Supplementary Videos 5,6). The tAB\(^\prime\) polytype is quite similar structurally to a twisted bilayer considering that both the large and small moiré patterns in the tAB\(^\prime\) trilayer relax into triangular domains.\supercite{yoo2019atomic,kazmierczak2021strain,craig2023local} With this in mind, we analyze the field-dependent structural distortions in both moiré lengthscales of the tAB\(^\prime\) structure and compare them to analogous twisted \(\mathrm{WSe_2}\) bilayers in Fig. 4b (twisted bilayer images provided in Supplementary Video 7). We do not observe marked differences in behavior between the smaller moiré in the trilayer (\(\theta_{23} = 0.25\degree\)) and a twisted bilayer with a similar interlayer rotation (\(\theta = 0.31\degree\)). Interestingly, the larger moiré in the trilayer (\(\theta_{12}<0.05\degree\)), has a considerably lower polarizability than its twisted bilayer counterpart (\(\theta\approx 0.03\degree\)).

DF-TEM images of the tAB\(^\prime\) structure show that the domain walls of the larger moiré are roughly commensurate with the domain walls in the smaller moiré, and as a field is applied, these two sets of domain walls appear to be pinned to one another (Fig. 4c). At one point, around -0.01 V/nm, the domain wall of the large moiré de-pins from that of the small moiré and jumps to an adjacent domain wall in the small moiré, leading to the S-shaped polarization curve in Fig. 4b. To quantitatively analyze this pinning effect, we track how the side lengths (\(\lambda\)) of the triangular domains in the small moiré (from Fig. 4c) evolve as a function of field along the length of one of the large moiré domain walls over three cycles of biasing. The results illustrate that the small moiré domains stretch anisotropically as the large moiré deforms in response to the field, particularly at positive field values (Fig. 4d). This increase in the periodicity of the small moiré could be facilitated by either a reduction in the local interlayer rotation in these domains or by a reduction in local lattice mismatch. Considering that one or more layers are sliding and stretching in response to the applied field, it is unlikely that the local lattice mismatch between layers is decreasing (we would expect the converse instead), and therefore we conclude that changes in local rotational offset between the middle and top layers (\(\theta_{23}\), Fig. 4e) drive the observed deformation of the small moiré domains. Fig. 4f,g shows how the average values of \(\lambda\) and \(\theta_{23}\) evolve for a group of domains from the region in Fig. 4c over three biasing cycles (field profile in Fig. 4h), showing that the small moiré length increases and the local twist angle decreases by up to factor of two as a result of pinning between the domain walls across the two interfaces. Importantly, these results indicate that the interfaces in polar multilayer heterostructures are not decoupled from one another and cooperative effects can influence observed switching dynamics, such as the reduction in polarizability observed in Fig. 4b. Notably, the results in Fig. 4c–e also demonstrate an electrical route to controlling superlattice wavelength in moiré structures.

%%%%%%%%%%%%%%%%%%%%%%%%%%%%%%%%%%%%%%%%%%%%%%%%
%% FIGURE 5
%%%%%%%%%%%%%%%%%%%%%%%%%%%%%%%%%%%%%%%%%%%%%%%%
\begin{figure*}[htbp]
    \centerline{\includegraphics[width=90mm]{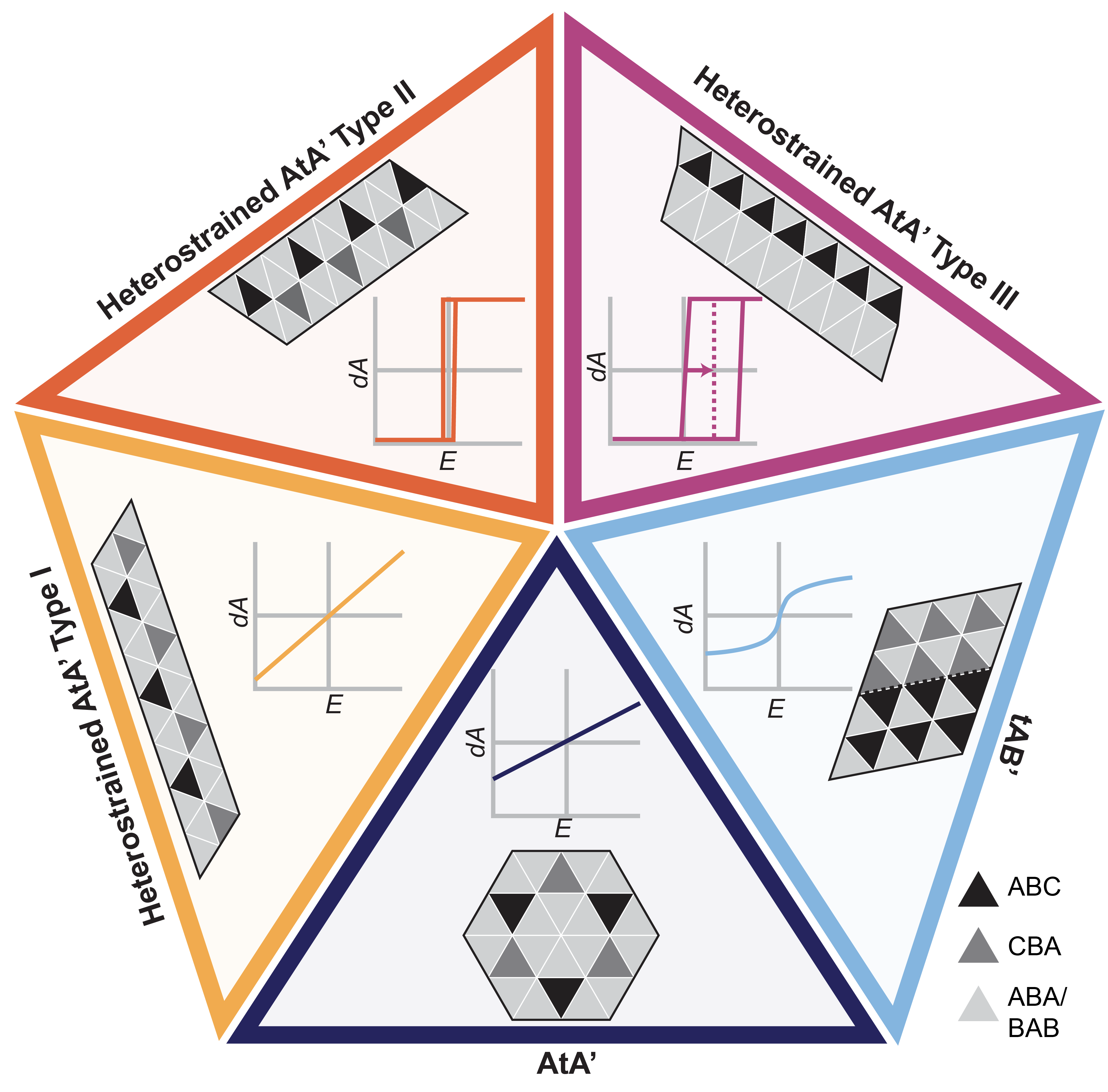}}
    \caption{\textbf{Design principles for tuning polarization switching dynamics in twisted \(\mathbf{WSe_2}\) trilayers.} Summary of the various types of polar domain structures studied in this work and their respective switching responses.}
\end{figure*}

\section*{Conclusions}

Two-dimensional vdW materials have exhibited promise as FE components in emerging technologies\supercite{wu2021two,wang2023towards} due to their robustness to depolarization at atomic thicknesses,\supercite{chang2016discovery,liu2016room,ding2017prediction,cui2018intercorrelated, fei2018ferroelectric,yuan2019room,higashitarumizu2020purely} their compatibility with silicon-based device schemes, and their ability to be stacked into multi-component heterostructures with sought-after functionalities, such as FE field-effect transistors (FE-FETs)\supercite{huang2020gate} and multiferroic devices.\supercite{gong2019multiferroicity,dou2022two}
Here, we have revealed that a diverse range of polarization switching dynamics can be accessed in twisted trilayer heterostructures due to the presence of more than one moiré periodicity as well as interactions between all three layers, as summarized in Fig. 5. Namely, changing relative rotations between layers and intralayer uniaxial strain fields leads to a variety of polar domain structures with different polarizabilities, coercivities, and intrinsic thermodynamic biases. This work paves the way for engineering polarization switching and structural dynamics in polar multilayer systems, particularly as developments in 2D heterostructure fabrication and strain engineering continue to advance. For example, variations in the arrangements of polar domains in the AtA\(^\prime\) versus tAB\(^\prime\) structures as well as the highly localized nature of preferential polarization states in the AtA\(^\prime\) polytype opens avenues for the design of structures and surfaces with deterministically patterned polarization. Even in untwisted polar vdW multilayers, manipulation of intralayer strain could afford control over consequent switching dynamics. In moiré multilayers comprised of semiconducting vdW materials, including \(\mathrm{WSe_2}\), such variability in polar order and switching responses could also lead to the emergence of exotic, electrically-tunable moiré exciton responses, beyond those that have been previously observed in bilayer heterostructures.\supercite{huang2022excitons}   

\newpage
\section*{Methods}
\subsection*{Sample preparation}
Monolayer \(\mathrm{WSe_2}\) (HQ Graphene), 2–5 layer graphite (Graphene Supermarket Kish graphite) and \(<15\) nm-thick hBN (grown by collaborators T.T. and K.W.) were mechanically exfoliated onto \(\mathrm{SiO_2}\)/Si substrates and selected based on optical contrast and atomic force microscopy measurements. Heterostructures were fabricated using the cut-and-stack dry transfer method.\supercite{kim2016van} A tungsten scanning tunneling microscope (STM) tip was first used to cut each \(\mathrm{WSe_2}\) monolayer into three pieces. A polybisphenol-A-carbonate/polydimethylsiloxane (PC/PDMS) stamp was then used to pick up the top graphene electrode followed by the top hBN and the first third of a pre-cut \(\mathrm{WSe_2}\) monolayer. The remaining \(\mathrm{WSe_2}\) portions were then sequentially rotated and picked up to construct the desired trilayer structure, followed by pick-up of the bottom hBN and finally the bottom graphite. We note that Device 2 (Supplementary Table 1) had hBN only on the top but shows similar behavior to the other devices which had hBN on both sides of the \(\mathrm{WSe_2}\). The final stacks were stamped onto Protochips electrical e-chips with 5 \(\mu\)m wide holes and Au pre-patterned electrodes in a four-point configuration and were then annealed under vacuum at 350\(\degree\)C to improve adhesion to the substrate. Lastly, custom electrical contacts were made from the pre-patterned electrodes to the top and bottom graphite sheets using e-beam lithography followed by reactive ion etching and thermal evaporation of Cr/Au (2 nm/100 nm). Further details and images of devices are provided in Supplementary Section 1.
 
\subsection*{Electron microscopy imaging}
Electron microscopy was performed at the National Center for Electron Microscopy in the Molecular Foundry at Lawrence Berkeley National Laboratory. Dark-field TEM images were collected using a Gatan UltraScan 1000 camera on a Thermo Fisher Scientific Titan-class microscope operated at 60 kV. Three frames each with an acquisition time of 5 s were summed to produce each dark-field image. A bias voltage was applied between top and bottom graphite electrodes of each device during imaging using a Protochips Aduro double-tilt biasing holder connected to a Keithley 2650 sourcemeter. Selected area electron diffraction patterns were collected on the same microscope at 60 kV using a Gatan Orius 830 camera, summing 16 frames with an exposure time of 0.1 s each.

Four-dimensional STEM datasets were acquired using a Gatan K3 direct detection camera located at the end of a Gatan Continuum imaging filter on a TEAM I microscope (aberration-corrected Thermo Fisher Scientific Titan 80–300). The microscope was operated in energy-filtered STEM mode at 80 kV with a 10 eV energy filter centred around the zero-loss peak (to reduce background from inelastic scattering), an indicated convergence angle of 1.71 mrad, and a beam current of 40–50 pA. These conditions yield an effective probe size of 1.25 nm (full-width at half-maximum value). Diffraction patterns were collected using a step size of 2 nm with 200 x 200 beam positions, covering an area of 400 nm x 400 nm. The K3 camera was used in full-frame electron counting mode with a binning of 4 × 4 pixels in each diffraction pattern and an energy-filtered STEM camera length of 800 mm. Diffraction patterns were acquired with an exposure time of 13 ms per pattern, which is the sum of multiple counted frames.

\subsection*{4D-STEM data analysis}
Analysis of 4D-STEM data was performed using Python on a personal computer. Published modules\supercite{savitzky2021py4dstem} were used for Bragg disk detection and integration to yield virtual dark-field images (Supplementary Fig. 5). All other code was custom written by the authors. Multislice simulations were carried out using the ABTEM\supercite{abtem} software package. Additional details are provided in Supplementary Section 3.

\newpage
\printbibliography

\section*{Data Availability}
All dark-field TEM images supporting the findings of this study are contained within the main text and supplementary files. Raw 4D-STEM data sets will be publicly accessible on Zenodo upon publication of the manuscript.

\section*{Code Availability}
The computer code for generation of colored stacking order maps and multislice simulations is available from the authors upon request.

\section*{Acknowledgements}
This work was supported by the U.S. National Science Foundation (NSF) under award no. DMR-2238196 (D.K.B.). M.V. acknowledges support from a University of California, Berkeley Philomathia Graduate Fellowship. I.M.C. acknowledges support from a University of California, Berkeley Berkeley Fellowship and a National Defense Science and Engineering Graduate (NDSEG) Fellowship under contract FA9550-21-F-0003 sponsored by the Air Force Research Laboratory (AFRL), the Office of Naval Research (ONR) and the Army Research Office (ARO). Work at the Molecular Foundry, LBNL was supported by the Office of Science, Office of Basic Energy Sciences, the U.S. DOE under Contract no. DE-AC02-05CH11231. Other instrumentation used in this work was supported by grants from the Gordon and Betty Moore Foundation EPiQS Initiative (Award no. 10637, D.K.B.), Canadian Institute for Advanced Research (CIFAR–Azrieli Global Scholar, Award no. GS21-011, D.K.B.), and the 3M Foundation through the 3M Non-Tenured Faculty Award (no. 67507585, D.K.B.). K.W. and T.T. acknowledge support from the Elemental Strategy Initiative conducted by the Ministry of Education, Culture, Sports, Science and Technology, Japan (grant no. JPMXP0112101001) and Japan Society for the Promotion of Science, Grants-in-Aid for Scientific Research (KAKENHI; grant nos. 19H05790, 20H00354 and 21H05233).

\section*{Author Contributions}
M.V.W. and D.K.B. conceived the study. M.V.W., N.D., and N.K. designed and fabricated the samples. M.V.W. and R.D. acquired TEM and 4D-STEM data. M.I. performed multislice simulations with input from I.M.C.. I.M.C. wrote the code used for generation of color-coded virtual dark-field images. T.T. and K.W. provided the bulk hBN crystals. M.V.W. processed and analyzed the data. M.V.W. and D.K.B. wrote the manuscript with input from all co-authors.

\section*{Competing Interests}
The authors declare no competing interests.

\section*{Additional Information}
Correspondence and requests for materials should be emailed to D.K.B. \\(email: bediako@berkeley.edu).

\end{document}